\documentclass[draft]{agujournal2019}
\usepackage{url} %this package should fix any errors with URLs in refs.
\usepackage{lineno}
\usepackage{soul}
\draftfalse

\journalname{Geophysical Research Letters}

\usepackage{xspace}
\newcommand{\kms}{km~s$^{-1}$\xspace}
\newcommand{\degrees}[1]{#1$ ^{\circ} $\xspace}

\usepackage{verbatim}

\begin{document}

%% ------------------------------------------------------------------------ %%
%
%  Title
%
%% ------------------------------------------------------------------------ %%
\title{Complex crater formation by oblique impacts on the Earth and Moon}

%% ------------------------------------------------------------------------ %%
%
%  AUTHORS AND AFFILIATIONS
%
%% ------------------------------------------------------------------------ %%
\authors{T. M. Davison\affil{1}, G. S. Collins\affil{1}}

\affiliation{1}{Impacts \& Astromaterials Research Centre, Department of Earth Science and Engineering, Imperial College London, London SW7 2AZ, U.K.}

\correspondingauthor{Thomas Davison}{thomas.davison@imperial.ac.uk}

\begin{keypoints}
\item We have simulated the formation of complex impact craters at a range of impact angles and velocities
\item Complex craters formed at oblique angles are larger than existing scaling laws predict
\item The overall effect of impact obliquity on crater populations is not as strong as earlier calculations suggested
\end{keypoints}

%% ------------------------------------------------------------------------ %%
%
%  ABSTRACT and PLAIN LANGUAGE SUMMARY
%
%% ------------------------------------------------------------------------ %%

\begin{abstract}
Almost all meteorite impacts occur at oblique incidence angles, but the effect of impact angle on crater size is not well understood, especially for large craters. To improve oblique impact crater scaling, we present a suite of simulations of complex crater formation on Earth and the Moon over a range of impact angles, velocities and impactor sizes. We show that crater diameter is larger than predicted by existing scaling relationships for oblique impacts and for impacts steeper than \degrees{45} shows little dependence on obliquity. Crater depth, volume and diameter depend on impact angle in different ways such that relatively shallower craters are formed by more oblique impacts. Our simulation results have implications for how crater populations are determined from impactor populations and vice-versa. Our results suggest that existing approaches to account for impact obliquity may underestimate the number of craters larger than a given size by as much as 40\%.
\end{abstract}

\section*{Plain Language Summary}
The relationship between impact crater size and impactor properties, such as size and speed, is key to comparing impactor and crater populations on different planets and dating planetary surfaces. 
Most of our understanding of this relationship, however, comes from numerical simulations of vertical-incidence impacts, and laboratory impact experiments at relatively low speed, which are comparatively rare in nature. 
Here we present results of numerical simulations of large crater formation on Earth and the Moon, for a range of oblique impact angles and speeds more typical of planetary scale impacts. 
We find that while crater size decreases as the impact angle becomes shallower, crater diameter, depth and volume are all affected by impact angle in different ways. 
Most importantly, we find that crater diameter depends less on impact angle than previously thought, especially for steeply inclined impacts. This implies that typical asteroid impacts on planetary surfaces form larger craters, and large craters are formed more frequently, than is currently assumed.

%% ------------------------------------------------------------------------ %%
%
%  TEXT
%
%% ------------------------------------------------------------------------ %%

\section{Introduction}

Impact craters are ubiquitous features on all solid solar system bodies. The final morphology of a crater depends on its size: small craters have simple, bowl-shaped cavities, while larger craters are subject to more profound late-stage, gravity-driven collapse, leading to complex morphologies including flat crater floors, central peaks or rings of peaks, and a terrace zone near the crater rim \cite<e.g.>{grieve_constraints_1981, melosh_impact_1989, french_traces_1998}. 
Due to the large length scales necessary to produce these characteristic features of complex craters, they are not reproducible in laboratory experiments. 
Crater size depends principally on impactor properties, for example the mass, velocity and impact angle \cite{holsapple_scaling_1986, schmidt_recent_1987}. 
Understanding this link between impactor properties and crater size is useful, as it allows estimates of the size and speed of the impactor to be inferred from an observed crater \cite<e.g.>{ivanov_numerical_2002, ivanov_numerical_2005, johnson_spherule_2016, collins_steeply-inclined_2020}. It can also be used to predict a population of craters on the surface of a planet or asteroid of a given age \cite<e.g.>{marchi_onset_2012, marchi_missing_2016, bottke_late_2017}. 

Scaling laws have been developed previously to link impactor and target properties to crater size \cite<e.g., >{holsapple_scaling_1986, schmidt_recent_1987, holsapple_scaling_1993}. These have typically relied on vertical-incidence laboratory experiments and numerical models. However, the majority of impacts occur at an oblique incidence angle: \degrees{45} is the most probable impact angle, and $\sim~90\%$ of all impacts occur at angles $<~70^\circ$ from the target plane \cite{shoemaker_interpretation_1962}. %
Previous laboratory experiments \cite{gault_experimental_1978, burchell_crater_1998} and numerical models \cite{elbeshausen_scaling_2009, davison_numerical_2011} have investigated the effects of impact angle on crater size; however, these studies have tended to use a low impact velocity ($<7$~\kms), and to simulate small or simple craters where growth is dominated by target strength and/or exhibit little or no late-stage, gravity-driven modification of the crater rim or floor. Thus, doubt remains over how to include impact angle in crater scaling, especially for larger, complex craters where the crater undergoes substantial collapse \cite{johnson_spherule_2016}. In low-velocity impacts, crater dimensions decrease as the impact becomes more oblique (i.e., as impact angle to the horizontal decreases) if all other impact parameters are held constant \cite{gault_experimental_1978, burchell_crater_1998}. This suggests that as impact trajectory becomes shallower the coupling of impactor momentum and energy with the target becomes less efficient. A widely adopted approach to describe this reduction in coupling efficiency is to assume that only the vertical component of the impactor momentum (and energy) contributes to crater growth and the horizontal component makes no contribution \cite{chapman_cratering_1986, elbeshausen_scaling_2009, davison_numerical_2011}. However, \citeA{ivanov_numerical_2002} noted that for higher-velocity (20~\kms) impact simulations, the difference in crater size between the vertical and oblique incidence impacts was less pronounced. They hypothesised that, above a critical angle, a different coupling regime may exist at high impact speeds, in which the coupling of momentum and energy is equally efficient, regardless of impact angle. 

To explore cratering efficiency in large-scale, high-speed oblique impacts, we simulate complex crater formation using the iSALE3D shock physics code. We extend previous work to higher impacts speeds (up to 30~\kms), to probe potentially different regimes of cratering \cite{ivanov_numerical_2002}, and to larger scales on Earth and the Moon (up to $\sim$200~km), where the excavated transient crater undergoes major gravity-driven collapse of the crater rim and uplift of the crater floor to form a complex crater. We show that crater depth and diameter depend on impact angle in different ways. While crater depth depends only on the vertical component of the impactor momentum (and energy), crater diameter (and volume) depend(s) on both vertical and horizontal components. Using these results, we quantify the effect that impact obliquity has on crater populations.

\section{Methods}
The iSALE3D shock physics code \cite{collins_isale-dellen_2016, elbeshausen_scaling_2009, elbeshausen_isale-3d:_2011} was used to simulate oblique incidence complex cratering events. The code uses a solver as described in \citeA{hirt_arbitrary_1974} and \citeA{amsden_sale-3d:_1981}. iSALE3D has previously been used to investigate the effects of impact angle on crater formation: \citeA{elbeshausen_scaling_2009} simulated low-velocity (7~\kms) impacts into cohesionless materials, and found an angle dependence of crater size consistent with experiments in granular materials \cite<e.g.>{gault_experimental_1978}; \citeA{davison_numerical_2011} simulated impacts into strong metal targets that very closely replicated experimental impacts \cite{burchell_crater_1998} including key dependencies between crater dimensions and impact angle. Here we extend these investigations to the gravity-dominated regime of complex cratering.

The target was composed of a 33~km thick granitic crust overlying a dunite mantle; a similar model set up to recent simulations of the Chicxulub impact \cite{collins_steeply-inclined_2020}. Both materials used an ANEOS-derived tabular equation of state \cite{pierazzo_reevaluation_1997, benz_origin_1989}.  The impacting asteroid was modelled as a sphere, and also used the ANEOS for granite. This material was chosen to match the crust, because of a current limitation of iSALE3D which does not allow more than one boundary between materials in a grid cell. 

iSALE3D includes a strength model appropriate for geologic materials \cite{collins_modeling_2004, melosh_dynamic_1992, ivanov_implementation_1997}. Strength parameters for the crust and mantle were chosen based on previous simulations of the Chicxulub impact, using iSALE3D \cite{collins_steeply-inclined_2020}, iSALE2D \cite{collins_dynamic_2008, christeson_mantle_2009, morgan_formation_2016} and SALEB \cite{ivanov_numerical_2002, stoffler_origin_2004}, and can be found in Table S1 in the supporting material. The acoustic fluidization ``block model'' \cite{wunnemann_numerical_2003} was used to allow late-stage collapse of the transient crater, with parameters also based on  \citeA{collins_steeply-inclined_2020}. 

To explore the effect of surface gravity, two sets of simulations were performed with gravity appropriate for the Earth or Moon. 
Impact velocity ($u$) was varied between 10--30 \kms on Earth, and 5--30 \kms on the Moon. Impactors had diameters ($L$) of 6.22, 8.96 and 14~km. 
We used a range of impact angles ($\theta$, defined as the angle between the impactor trajectory and the horizontal target surface) between \degrees{30} and \degrees{90} (vertical incidence). All models employed a spatial resolution of 14~cppr (cells per projectile radius), which was previously found to reproduce the Chicxulub crater on Earth well \cite{collins_steeply-inclined_2020}. Technical details of the model setup can be found in the Supporting Information. 

Formation of complex craters involves two major phases of crater growth. During initial excavation, a deep, bowl-shaped crater grows by outward displacement and ejection. The form of the crater at the end of excavation is termed the transient crater. Subsequently, the steep walls of the crater rim collapse inwards while the floor of the crater is uplifted by gravitational instability. During this modification stage the crater depth and volume decrease, while diameter increases. Here we follow the common convention to measure transient crater dimensions at the time of maximum crater volume \cite{kenkmann_impact_2005, elbeshausen_scaling_2009}. 
Transient crater diameter is measured at the pre-impact surface; crater depth is measured from the pre-impact surface; and transient crater volume is the volume of the crater excavated below the pre-impact surface. 
Since the along-range and cross-range diameters can sometimes be quite different in oblique impacts, here we use the diameter of a circle with an area equivalent to the area of the crater planform at the pre-impact elevation to allow direct comparisons between models (termed the ``equivalent diameter'', $D_{tr,eq}$). 

Final crater measurements were taken after all material in the simulation had ceased moving (with the exception of some melt). To find the final rim-to-rim diameter ($D_{f,rr}$), radial surface profiles were drawn at azimuths with a spacing of \degrees{2}, from the centre of the crater measured at the pre-impact surface. For each profile, the maximum height was found; the outline of the crater rim was found by connecting these peaks. The reported rim-to-rim diameter is the radius of a circle with an area equivalent to the area of the crater rim outline.

\section{Results}

\begin{figure}
  \noindent\includegraphics[width=160mm]{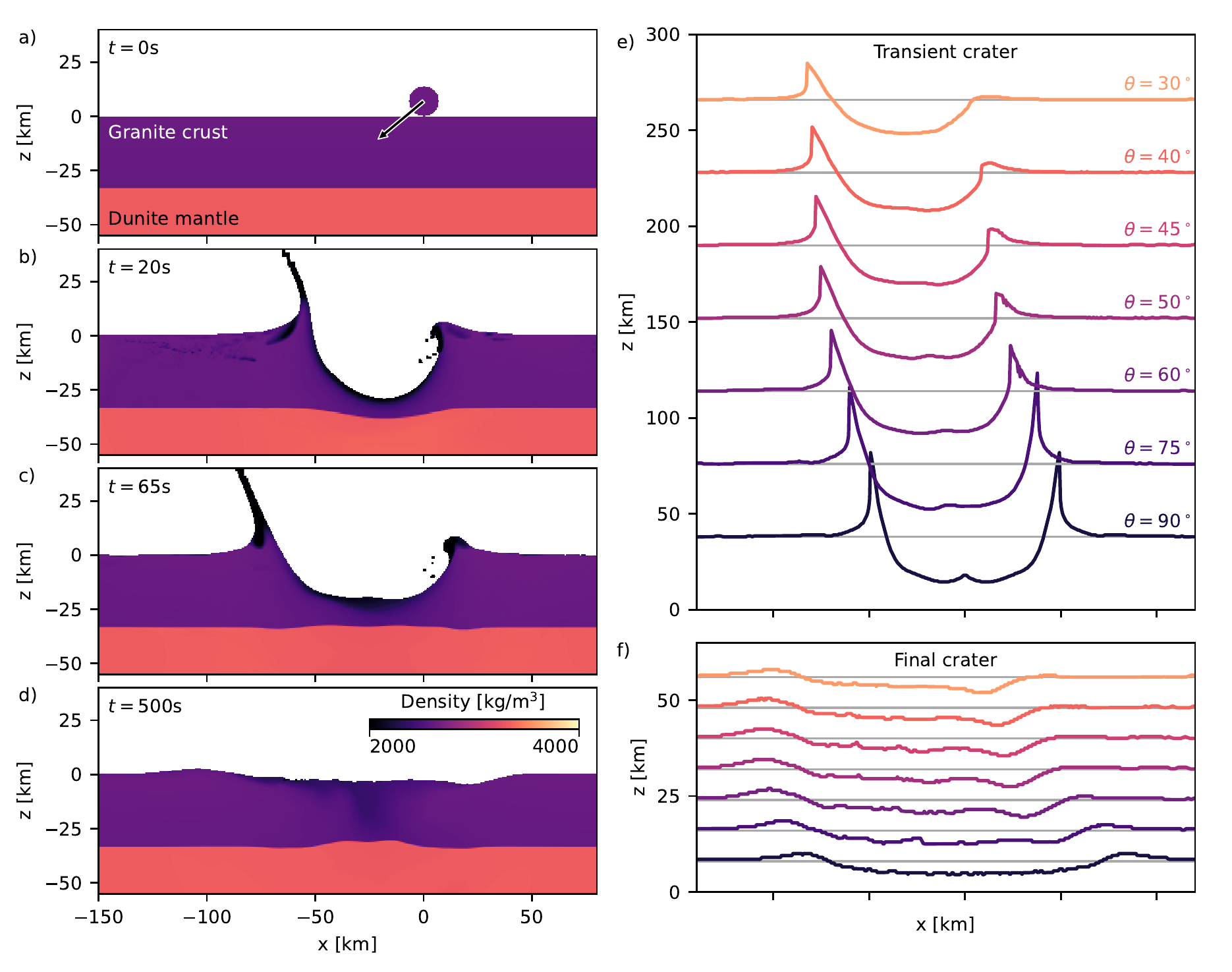}
    \caption{Snapshots from a simulation with $L=14$~km, $u=20$~\kms and $\theta=45^\circ$ on Earth. The frames are slices along the symmetry plane of the simulation, and show (a) the initial conditions, (b) the crater at its maximum depth (c) maximum volume and (d) final morphology. The arrow in (a) shows the projectile's trajectory. Transient (e) and final (f) crater surface profiles taken along the symmetry plane, for simulations with $L=14$~km, $u=20$~\kms on Earth, for impact angles \degrees{30}--\degrees{90}. Note in (f), the $z$-axis scale is exaggerated by a factor of 2.}
\label{Fig:frames}
\end{figure}

Transient craters ranged in diameter from 38--98~km on Earth and 54--150~km on the Moon. Final craters ranged in diameter from 72--200~km on Earth and 80--230~km on the Moon (see Tables S2 \& S3 in the Supporting Information). The craters exhibit flat crater floors and a broad central structural uplift that extends to the base of the crust (Figure~\ref{Fig:frames}). A detailed analysis of final crater morphologies is not possible due to the spatial resolution of our simulations; however, as in previous simulations of complex crater formation \cite{collins_steeply-inclined_2020}, in all cases final crater formation involves the formation of an over-heightened central uplift and its subsequent collapse, resulting in a central peak or peak ring final crater morphology.
The initial geometry of a typical simulation is shown in Figure~\ref{Fig:frames}a, for $L=14$~km, $v=20$~\kms and $\theta=45^\circ$ on Earth. The crater reaches its maximum depth at $t=20$~s, (Figure~\ref{Fig:frames}b), and its maximum volume at $t=65$~s (Figure~\ref{Fig:frames}c).  
Note that the start of the collapse process is not the same at all azimuths in oblique impacts---in this simulation, collapse on the uprange side begins 5--10 seconds before the time of maximum volume, and downrange collapse starts 10--15 seconds after the time of maximum volume. The amount of enlargement in the uprange direction (i.e., the distance the crater wall moved outwards during collapse) is approximately 20\% greater than in the downrange direction. By 500~s, modification has ended and the crater has reached its final morphology (Figure~\ref{Fig:frames}d). 

\subsection{Dependence of Crater Morphology on Impact Angle}

Surface profiles of the transient and final craters (along the symmetry plane) reveal several morphological differences that are exhibited as the impact angle changes from vertical (\degrees{90}) to oblique (\degrees{30}) incidence (Fig.~\ref{Fig:frames}e--f). Both transient and final craters are centred further downrange from the impact point for more oblique impacts. The transient crater depth/diameter ratio decreases with increasing obliquity for all suites of simulations, consistent with impact experiments in quartz sand \cite{gault_experimental_1978}. 

The slope of the downrange wall of the transient crater is less steep in more oblique impacts (Fig.~\ref{Fig:frames}e). This has previously been observed in modelling \cite<e.g.>{elbeshausen_scaling_2009,davison_numerical_2011} and experimental studies \cite<e.g.>{gault_experimental_1978}. In most simulations, the steeper uprange wall leads to more enlargement of the crater on that side compared to the enlargement on the downrange side. The collapse of the uprange wall also starts earlier than the downrange collapse. In the uprange direction, the rim height of the final crater decreases in more oblique impacts (Fig.~\ref{Fig:frames}f); for the \degrees{30} impact, there is almost no detectable rim on the uprange side of the crater.

The floor of the final crater also exhibits asymmetries for impact angles $<60^\circ$. The crater floor on the the downrange side of the crater has a higher elevation compared to the uprange side (Fig.~\ref{Fig:frames}f), due to melt being distributed on the downrange side of the crater. While the crater walls cease moving by the simulation end time, melt can still be mobile. Thus, the trough on the uprange side of the crater at the end of our simulations will likely be infilled later by down-slope migration of melt.

\subsection{Dependence of Crater Dimensions on Impact Angle}
\label{Sect:Scaling}

\begin{figure}
  \noindent\includegraphics[width=95mm]{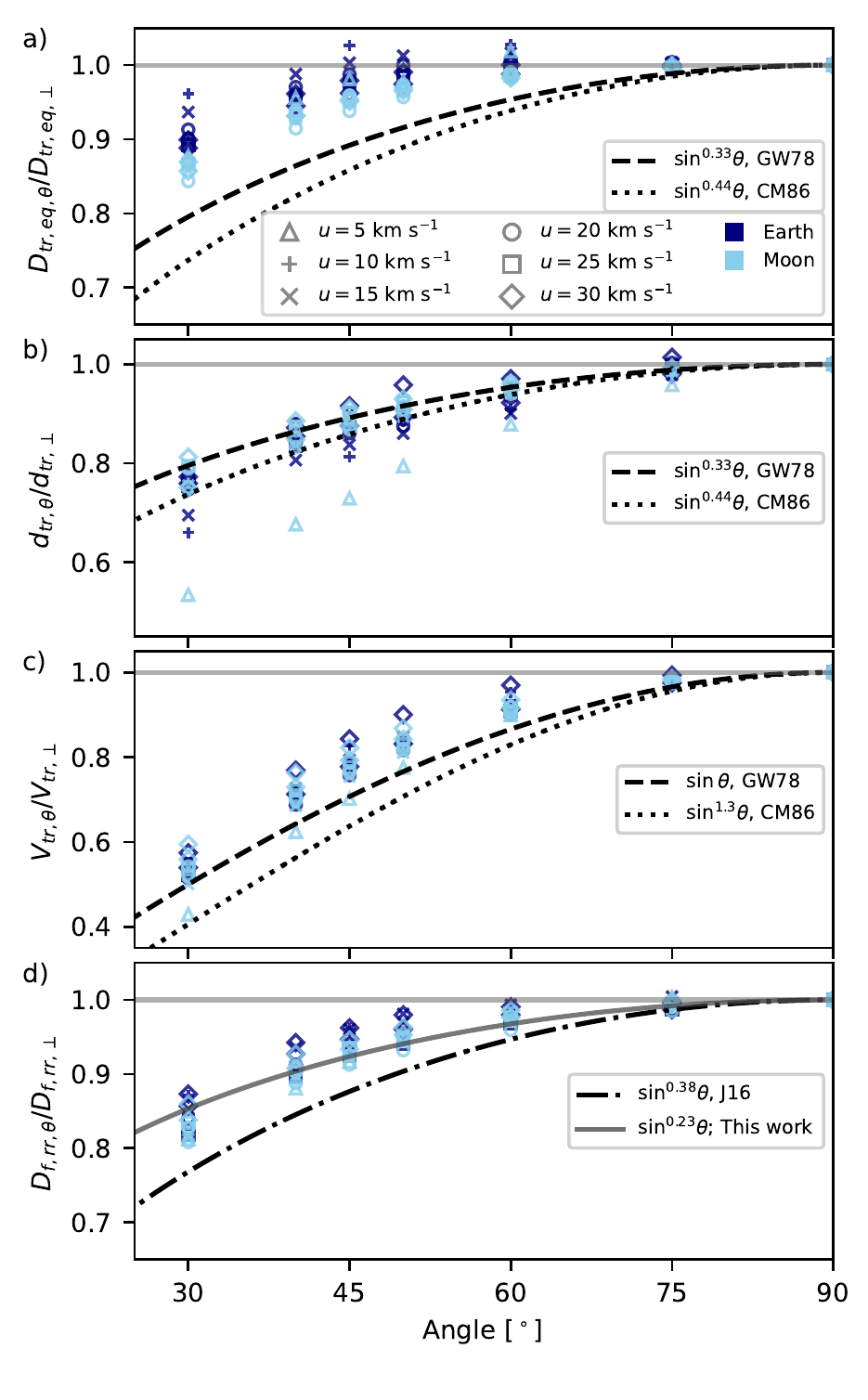}
    \caption{Crater measurements normalised by the equivalent vertical-incidence crater. (a) Transient diameter; (b) Transient depth; (c) Transient volume; (d) Final rim-to-rim diameter. The lines plotted correspond to different angle scaling from the literature: GW78: \citeA{gault_experimental_1978}, CM86: \citeA{chapman_cratering_1986}, J16: \citeA{johnson_spherule_2016}. The solid grey line in (d) is a fit to the final rim-to-rim diameters from this work.
    }
\label{Fig:anglescaling}
\end{figure}

Figure~\ref{Fig:anglescaling} shows how different crater dimensions are affected by impact angle. Crater dimensions are normalised by those of the equivalent crater formed at vertical incidence. 
For impacts steeper than \degrees{45}, transient crater diameter shows very little dependence on impact angle (Figure~\ref{Fig:anglescaling}a). All transient diameters for angles \degrees{45} and steeper are within $\sim~5$\% of the vertical incidence simulation. 

Transient crater depth (Figure~\ref{Fig:anglescaling}b) and volume (Figure~\ref{Fig:anglescaling}c) both show a greater dependence on impact angle than crater diameter; all craters formed at angles shallower than \degrees{75} are smaller than the equivalent crater formed at vertical incidence. There is also some co-dependence of crater depth on impact velocity: for low velocity impacts (5~\kms on the Moon, and 10--15~\kms on Earth), the normalised depths of craters formed at shallow impact angles are smaller than the equivalent impacts at higher velocity. A larger effect of impact angle on crater depth at lower impact speeds is consistent with the simulation results of \citeA{ivanov_numerical_2002}. The influence of impact angle on crater diameter, depth and volume does not appear to be affected by the impactor size (in the range $L=6.22\textrm{--}14$~km) or target planet (gravity).

Final crater diameter (Figure~\ref{Fig:anglescaling}d) shows a greater dependence on impact angle than transient crater diameter. This demonstrates that the expansion of the crater rim during the modification stage depends on the impact angle: while transient crater diameter does not depend on impact angle above \degrees{45}, the final crater diameter does. This is because the enlargement of the crater rim during modification depends on the geometry of the transient crater. Deeper transient craters exhibit more enlargement, and the transient depth depends more on impact angle than does the diameter. 

\section{Oblique Impact Crater Scaling} 
\label{Sect:scaling_discussion}
The relationship between crater size and impactor and target parameters is complex \cite<e.g.>{melosh_impact_1989, holsapple_scaling_1993, schmidt_recent_1987}. Hence, empirical scaling relationships derived from laboratory experiments and numerical simulations are typically used \cite<known as $\pi$-group scaling;>{holsapple_scaling_1993}, in which dimensionless measures of impactor size ($\pi_2=1.61gL/u^2$, where $g$ is the surface gravity) and crater size ($\pi_D=D_{tr,eq}\left(\rho/m\right)^{1/3}$, where $\rho$ is the target density and $m$ is the impactor mass) are related to each other with a simple power-law:
\begin{equation}\label{eq:piDpi2}
    \pi_D = C_D \pi_2^{-\beta},
\end{equation}
where $C_D$ and $\beta$ are material-specific constants. Commonly adopted values for dense rock are $C_D = 1.6$, $\beta = 0.22$ \cite{schmidt_recent_1987, melosh_impact_1989}. 

The influence of impact angle on $\pi$-group scaling is not well understood. A common method to quantify how impact angle influences crater size is to examine how a given crater dimension (e.g. diameter, $D$, depth, $d$, or volume, $V$) scales with impact angle, when normalised by the same dimension in a vertical-incidence impact event; for example $D_\theta / D_\perp = \sin^{\alpha_D}\theta$, or $V_\theta / V_\perp = \sin^{\alpha_V}\theta$, where $\alpha_D$ and $\alpha_V$ are constants to be determined. It is often assumed that $\alpha_V = 3\alpha_D$ which would imply that the crater depth-to-diameter ratio is independent of impact angle.
Pioneering oblique impact experiments showed that $\alpha_V=1$ for gravity-dominated impacts in cohesionless sand and $\alpha_V \approx 2$ for strength-dominated impacts in granite \cite{gault_experimental_1978}. Based on these trends, \citeA{chapman_cratering_1986} proposed that $\pi$-group scaling of crater dimensions could be extended to include angle dependence by replacing the impact speed $u$ with the vertical component of impact velocity $u\sin\theta$ in the definition of $\pi_2$. This simple and elegant approach, which assumes that the horizontal component of the impactor momentum (and energy) makes no contribution to crater size, gives $\alpha_V = 1$ for cohesionless sand ($\beta = 0.16\textrm{--}0.17$), and $\alpha_V = 1.7$ for dense rock or metal in the strength regime \cite{chapman_cratering_1986}. This form of scaling has been widely adopted \cite<e.g.>{ivanov_marsmoon_2001, marov_size-frequency_2001} because it is broadly consistent with results of low-velocity impact experiments \cite{gault_experimental_1978, burchell_crater_1998} and numerical simulations \cite{elbeshausen_scaling_2009, davison_numerical_2011}. Using this scaling and adopting a nominal value of $\beta=0.22$ in Equation~\ref{eq:piDpi2} for a dense rock target \cite{schmidt_recent_1987} implies $\alpha_V=1.32$ ($\alpha_D=0.44$).

On the other hand, simulations of low-velocity, gravity-dominated oblique impacts in dense, cohesionless targets offer more equivocal support for scaling by the vertical component of the impact velocity \cite{elbeshausen_scaling_2009}. Results from these simulations instead suggest $\alpha_V \approx 1$ ($\alpha_D \approx 0.33$) for a wide range of target friction coefficients. As a result, an alternative form of angle dependence in popular use is to simply scale the crater volume by $\sin\theta$ and diameter (or depth) by $\sin^{0.33}\theta$ \cite<e.g., >{collins_earth_2005, johnson_spherule_2016}. 

Our simulation results demonstrate that dependence of crater dimensions on impact angle is more complicated than either of these scaling approaches predict (Figure~\ref{Fig:anglescaling}a--c). Transient crater diameters formed at oblique incidence in this study are larger than those predicted using $\alpha_D=0.33$ (dashed line) or $\alpha_D=0.44$ (dotted line) (Figure~\ref{Fig:anglescaling}a). Transient crater volumes are also typically larger than predicted by $\alpha_V=1$ or $\alpha_V=1.32$ for $\theta\leq60^\circ$ (Figure~\ref{Fig:anglescaling}c). One suite of simulations (with $u=5$~\kms on the Moon) lies between those two scalings, but almost all other simulated craters have volumes larger than predicted by $\alpha_V=1$. On the other hand, transient crater depth is fit reasonably well by $\alpha_D=0.33$ or $\alpha_D=0.44$ for most simulations, particularly if the lowest impact speed simulations are excluded (Figure~\ref{Fig:anglescaling}b). 

Using $\alpha_D=0.33$, and assuming that the process of crater collapse is independent of impact angle, \citeA{johnson_spherule_2016} proposed that for complex craters final crater diameter should scale with $\alpha_D=0.38$. While agreement with this scaling is closer than for transient crater diameter, our oblique impact simulation results of final rim-to-rim crater diameters are also larger than predicted by \citeA{johnson_spherule_2016}. 

Our results demonstrate that different crater dimensions are affected by impact angle in different ways. A single scaling relationship is unable to describe the angle dependence of depth, diameter and volume. In particular, while crater depth appears to be independent of the horizontal component of the impact velocity, both crater diameter and volume depend to some extent on the impactor's horizontal motion. 

\citeA{ivanov_numerical_2002} hypothesised that the influence of impact angle on the transient crater size depends on impact velocity. They proposed the existence of two regimes: in low velocity oblique impacts, only the vertical component of the velocity affects crater size (i.e., $\alpha_D=0.44$) while in high velocity oblique impacts impact angle has no influence on crater size (i.e., $\alpha_D=0$) above a certain threshold angle. According to this idea, the threshold impact angle and speed between these regimes is defined by the vertical velocity necessary for the impactor to penetrate below the pre-impact surface during contact and compression; that is, before the impactor is unloaded from high pressure. High-speed impactors that quickly penetrate below the target surface are therefore able to couple all their momentum and energy with the target. 
We find partial support for this concept in our simulation results. Transient crater diameter appears to exhibit two regimes of angle dependence: at steep angles ($>45^\circ$) there is almost no dependence of diameter on impact angle ($\alpha_D=0$), whereas at shallow angles diameter drops abruptly. However, the transition angle between these regimes does not appear to depend on impact speed. On the other hand, the effect of impact angle on transient crater depth is diminished as impact speed increases, but there is no clear evidence of a distinct high-speed, angle-independent regime for depth. 

To account for impact angle in planetary crater scaling relationships, therefore, we propose that separate, empirical relationships based on our simulations should be used. For final crater diameter an angle exponent of $\alpha_D=0.23$ is suggested by our results, although further exploration of acoustic fluidization parameters would be prudent to confirm this trend. For transient crater diameter, a single fit to our simulation data gives a exponent of $\alpha_D=0.18$. 

\begin{figure}
  \noindent\includegraphics[width=95mm]{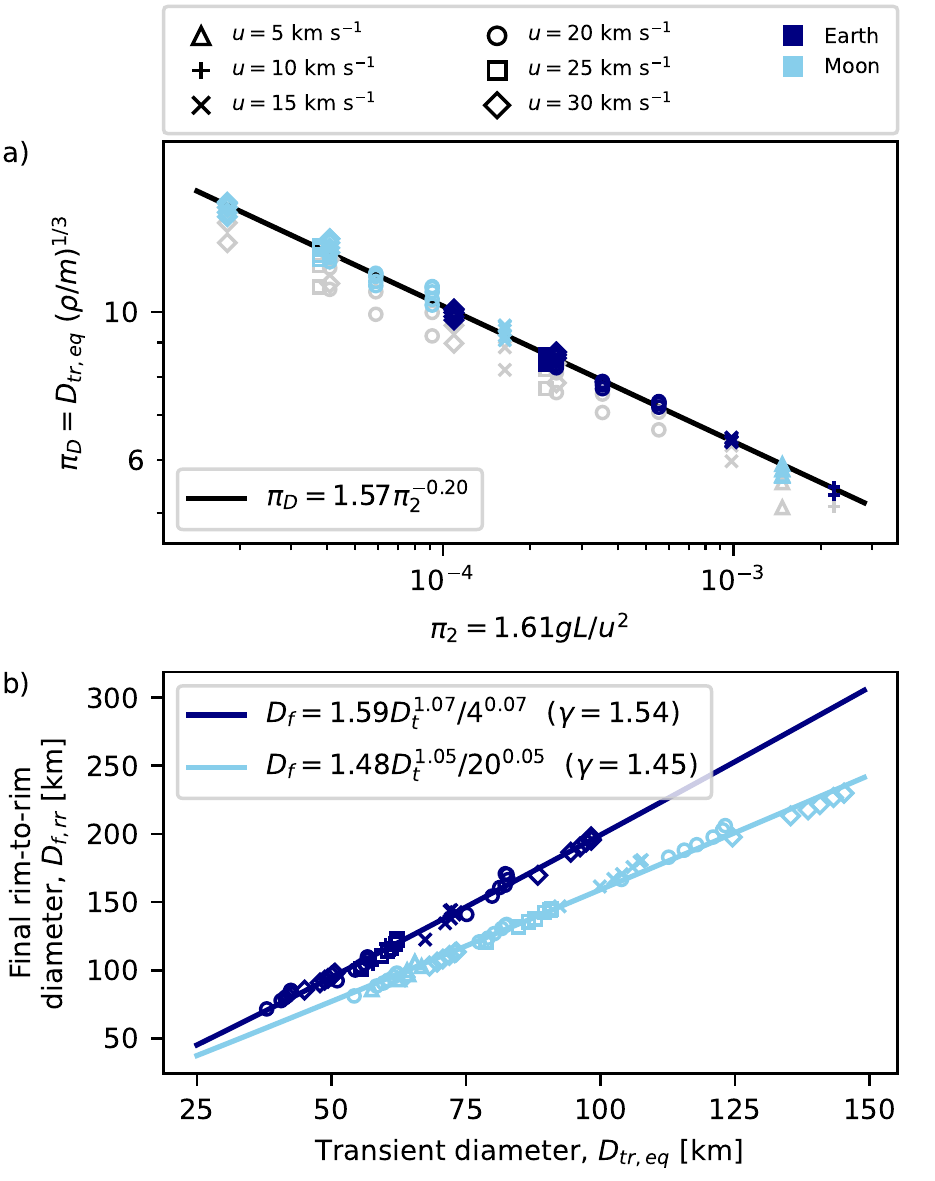}
    \caption{(a) All craters plotted in $\pi_D-\pi_2$ space. Impacts at $\theta \geq 45^\circ$ are colored light blue (Moon) and dark blue (Earth), while impacts more oblique than \degrees{45} are colored grey, and are not included in the least-squares fit (black line). (b) Final rim-to-rim diameter vs.\ transient crater diameter measured at the pre-impact surface.}
\label{Fig:piscaling}
\end{figure}

The observation that transient crater diameter is not affected significantly by impact angle for angles steeper than \degrees{45} is reinforced by plotting our results in  $\pi_2\textrm{--}\pi_D$ space (Equation~\ref{eq:piDpi2}; Fig.~\ref{Fig:piscaling}a). Impacts with angles shallower than \degrees{45} are shaded grey, while all others use light and dark blue colouring for the Moon and Earth, respectively. Fitting constants $C_D$ and $\beta$ in Equation~\ref{eq:piDpi2} for all impacts with $\theta \geq 45^\circ$ yields $C_D=1.57$ and $\beta=0.20$ ($R^2=0.99$), similar to values often adopted for dense rock \cite<$C_D=1.6, \beta=0.22$;>[]{schmidt_recent_1987, melosh_impact_1989}. 

Our results also provide insight into the enlargement of crater diameter caused by late-stage gravitational collapse of complex craters (Fig.~\ref{Fig:piscaling}b). While there is some dependence of the amount of crater rim enlargement on impact angle, our results support the use of a scaling relationship for crater enlargement as a function of crater size only. Final rim-to-rim crater diameter $D_{f,rr}$ is well described by:
\begin{equation} \label{Eq:tr2fin}
    D_{f,rr} = A {D_{SC}}^{-\eta}{D_{tr}}^{1+\eta}
\end{equation}
where $D_{tr}$ is the transient crater diameter at the pre-impact level, $D_{SC}$ is the final rim-to-rim diameter at the simple-to-complex transition, and $A$ and $\eta$ are constants \cite<e.g., >[]{johnson_spherule_2016}. Since the data collapse onto the power law given in Equation~\ref{Eq:tr2fin}, the effects of impact angle on this enlargement are shown to be minor compared to the influence of the transient crater size.

By assuming a value of $D_{SC}$ for the Earth and Moon, a least-squares fit can be used to estimate $A$ and $\eta$ (Fig.~\ref{Fig:piscaling}b).  On Earth, Brent crater \cite< $D_{f,rr}=3.8$~km; >{dence_terrestrial_1977} is one of the largest observed simple craters, which we use here to define the simple-to-complex transition of $D_{SC} = 4$~km. Fitting Equation~\ref{Eq:tr2fin}, we find $A=1.59, \eta=0.07$.
For the Moon, assuming $D_{SC} = 20$~km, we find $A=1.48, \eta=0.05$. These values of $\eta$ are close to the value of 0.08 proposed by \citeA{holsapple_scaling_1993}, but lower than that proposed by \citeA{johnson_spherule_2016} based on vertical impact simulations alone.

For simple craters, the ratio of final rim-to-rim diameter to transient crater diameter is a constant $\gamma = D_{f,rr}/D_{tr}$. To give a consistent relationship at the simple-to-complex transition, $A$ and $\eta$ are related to $\gamma$ by $\gamma=A^{\frac{1}{1+\eta}}$. Extrapolation of our simulation results to the simple-to-complex transition therefore implies simple crater enlargement ratios of $\gamma=1.54$ (Earth) and $\gamma=1.45$ (Moon). While this range of $\gamma$ is larger than often used values of 1.25--1.3 \cite{grieve_geometric_1984, collins_earth_2005, holsapple_scaling_1993}, it is consistent with recent simulations of both Barringer crater \cite{collins_numerical_2016} and Brent crater \cite{collins_numerical_2014} that well reproduce geological and geophysical observations.

\section{Implications for Crater Populations on Planetary Surfaces}
Our results have obvious implications for estimating impactor size for a specific event, such as the Chicxulub impact. The size of the impactor that triggered the K-Pg extinction has been controversial owing to conflicting interpretations of geophysical and geochemical evidence \cite{paquay_determining_2008, morgan_comment_2008}.
\citeA{collins_steeply-inclined_2020} showed that an impact angle of \degrees{45--60} is most consistent with geophysical observations of the Chicxulub crater. %
If we use the final crater scaling relationship from \citeA{johnson_spherule_2016} to predict the size of the Chicxulub impactor ($\alpha_D=0.38$), assuming the impactor and target densities are similar, and the impact happened at 20~\kms and at \degrees{45}, a 160~km diameter crater requires a 14~km diameter impactor. If, instead, we remove the angle dependence, as suggested by the transient crater scaling for angles $>45^\circ$ (Figure~\ref{Fig:anglescaling}a), the required impactor size reduces to around 12~km (a 14\% reduction in diameter and a 36\% reduction in mass).

Our revisions to oblique crater scaling also have implications for interpreting crater populations. \citeA{gault_experimental_1978} used the results of their experiments to show that the effects of obliquity would reduce the apparent population of craters on a planetary surface larger than a given diameter by a factor of $\sim 2$ when compared with a population formed by vertical impacts. In their derivation (their equations 1--8), they used kinetic energy scaling to relate impactor mass to crater size. Following a similar derivation using $\pi$-group scaling instead, the total number of craters larger than a given diameter formed at all trajectories $N_\theta$ can be expressed in terms of the number of craters formed if obliquity is neglected, $N_\perp$:

\begin{equation}\label{Eq:popfactor}
N_\theta = \left(\frac{2\left(1-\beta\right)}{2\left(1-\beta\right) - 3\alpha_D\psi}\right) N_\perp,
\end{equation}

\noindent where $\beta$ is the exponent from the $\pi_D\textrm{--}\pi_2$ scaling (Equation \ref{eq:piDpi2}), $\alpha_D$ is the constant which determines how much impact angle affects crater diameter, and $\psi$ is the exponent of the mass--number distribution of impactors ($N_i \propto m_i^\psi$). Since $\psi$ is a negative number, this implies that for any value of $\alpha_D > 0$, the effects of obliquity reduce the apparent number of craters. 

Using $\beta=0.22$ \cite{schmidt_recent_1987} we can then determine how different values of $\alpha_D$ affect the apparent number of craters. Here we use values of $\psi=-0.78$ \cite<for $10^{10}~\textrm {kg}<m_i<10^{16}~\textrm{kg}$;>{harris_population_2015} or $-0.93$ \cite<for $m_i<10^{10}~\rm{kg}$;>{bland_rate_2006}. Using the fit to final rim-to-rim craters ($\alpha_D=0.23$, Figure~\ref{Fig:anglescaling}d), we find that $N_\perp/N_\theta = 1.35\textrm{--}1.41$, and using the fit to transient crater diameters ($\alpha_D=0.18$, Figure~\ref{Fig:anglescaling}a) yields $N_\perp/N_\theta = 1.27\textrm{--}1.32$. Similar factors are obtained by numerical integrations using more complex fits to the dependence of crater diameter on impact angle observed in the simulation data.

These estimates of $N_\perp/N_\theta$ are substantially lower than 2 \cite{gault_experimental_1978}---our results therefore suggest that impact obliquity has less of an effect on crater populations than previously proposed. Rather than reducing the apparent population by a factor of 2, obliquity reduces the population by $\sim~30\textrm{--}40\%$. Conversely, the use of previously proposed crater size scaling relationships that account for impact angle \cite<e.g.>{ivanov_numerical_2002, johnson_spherule_2016} to convert an impactor population into a crater population likely underestimates the crater population size by 21--40\%.

The effect of impact angle on complex crater formation in the size range $\sim~70\textrm{--}230~\textrm{km}$ is less pronounced than previously thought, especially for hypervelocity ($>15$~\kms) impacts. Thus, the often-used simplification of vertical incidence in impact modelling is warranted, as a first approximation, for analysis of crater size in most impact scenarios on large planetary bodies. However, impact angle does influence crater size at low impact speeds and important asymmetries exist in the formation and final crater morphology of complex craters, particularly at angles shallower than 45 degrees. Future 3D oblique impact simulations will extend our understanding of the effect of impact angle on crater size and other structural dimensions to shallower angles, both smaller and larger impacts, as well as different planetary bodies (e.g., icy satellites).

\acknowledgments
We thank the developers of iSALE3D (isale-code.de), in particular Dirk Elbeshausen, Kai W\"unnemann, Jay Melosh and Boris Ivanov. TMD and GSC were supported by STFC grant ST/S000615/1. This work used the DiRAC Data Intensive service at Leicester, operated by University of Leicester IT Services, part of the STFC DiRAC HPC Facility (www.dirac.ac.uk). The equipment was funded by BEIS capital funding via STFC capital grants ST/K000373/1 and ST/R002363/1 and STFC DiRAC Operations grant ST/R001014/1. DiRAC is part of the National e-Infrastructure. 

\noindent {\bf Data Availability}: At present, the iSALE code is not fully open source. It is distributed via a private GitHub repository on a case-by-case basis to academic users in the impact community, strictly for non-commercial use. Scientists interested in using or developing iSALE should see http:// www.isale-code.de for a description of application requirements. Simulation input files, post-processing scripts and reduced output data files are available on Zenodo \cite{davison_dataset_2022} (https://doi.org/10.5281/zenodo.7041587).

%% ------------------------------------------------------------------------ %%
%% References and Citations

\bibliography{references}

\end{document}